\documentclass[aps, pra, twocolumn, superscriptaddress]{revtex4}
\usepackage{graphicx}
\usepackage{amsmath}
\usepackage{amssymb}

\usepackage[colorlinks, linkcolor=blue, urlcolor=magenta, citecolor=blue]{hyperref}

\begin{document}

\title{Bifurcations of nonlinear dynamics in coupled twin spin masers}

\author{Tishuo Wang}
\affiliation{Guangdong Provincial Key Laboratory of Quantum Metrology and Sensing, and School of Physics and Astronomy, Sun Yat-Sen University (Zhuhai Campus), Zhuhai 519082, China}
\affiliation{State Key Laboratory of Optoelectronic Materials and Technologies, Sun Yat-Sen University (Guangzhou Campus), Guangzhou 510275, China}

\author{Zhenhua Yu}
\email[]{huazhenyu2000@gmail.com}
\affiliation{Guangdong Provincial Key Laboratory of Quantum Metrology and Sensing, and School of Physics and Astronomy, Sun Yat-Sen University (Zhuhai Campus), Zhuhai 519082, China}
\affiliation{State Key Laboratory of Optoelectronic Materials and Technologies, Sun Yat-Sen University (Guangzhou Campus), Guangzhou 510275, China}

\begin{abstract} 
Spin masers are a prototype nonlinear dynamic system. They undergo a bifurcation at a critical amplification factor, transiting into a limit cycle phase characterized by a Larmor precession around the external bias magnetic field, thereby serving as a key frequency reference for precision measurements. Recently, a system of coupled twin spin masers placed in dual bias magnetic fields, involving simultaneously two intrinsic Larmor frequencies, has been studied~\cite{wang2023feedback}. Compared with previous spin masers, this setup exhibits new attractors such as quasi-periodic orbits and chaos in addition to the usual limit cycles and the trivial no signal fixed point. The richer dynamic phases imply the existence of bifurcations, whose nature has not been fully analyzed.
Here, to shed light on the nature of the bifurcations, we turn to a closely related system and 
systematically study the various bifurcations therein along different routes in parameter space. We identify the bifurcations as of the types including pitchfork, Hopf, homoclinic bifurcations, and saddle-node bifurcations of cycles. By both analytical and numerical methods, we reveal how various attractors interplay with each other and change their stability. We also quantitatively evaluate the locations where these bifurcations occur by tracking the both stable and even not easily detected unstable limit cycles. These findings deepen our understanding of the underlying mechanisms resulting in the rich dynamic phases in the coupled twin spin masers.

\end{abstract}

\maketitle

\section{Introduction}
 
 Bifurcation theory is a cornerstone of nonlinear dynamics, studying how a system changes qualitatively as a parameter varies~\cite{kuznetsov1998elements,seydel2009practical}.  It helps describe and predict transitions between different states or behaviors in complex systems~\cite{crawford1991introduction,guckenheimer1983nonlinear}. 
Because of the universality of nonlinearity in nature, bifurcations appear across a wide range of fields, including sociology, engineering, biology, chemistry, meteorology and physics etc~\cite{Strogatz2018Nonlinear}. 
   Bifurcations can manifest in various forms, including saddle-node, pitchfork, homoclinic, Hopf bifurcations, etc., each with unique characteristics and implications~\cite{guckenheimer1983nonlinear, Strogatz2018Nonlinear}.
 The occurrence of bifurcation is often accompanied by interplays between multiple attractors, such as "splitting", "collisions", or "coincidence", leading to the creation or annihilation of fixed points, limit cycles, quasi-periodic orbits, or chaos etc. Among them, the not easily detectable unstable limit cycles usually play a crucial role in leading to elusive phenomena~\cite{marsden2012hopf,drazin1992nonlinear, doedel2005numerical, Meijer2011}.
 For example, in the well-known Lorenz system, a Hopf bifurcation happens where a pair of fixed points become unstable by "absorbing" a pair of  gradually shrinking  unstable limit cycles as the Rayleigh number $r$ increases, leading the system to become entirely chaotic~\cite{lorenz1963deterministic,sparrow1982lorenz}.
 
 
 Advanced spin masers are representative nonlinear systems, which operate in a bias magnetic field while subject to an artificial feedback transverse magnetic field that generates an effective torque to balance spin relaxations~\cite{Inoue2016Frequency,Yoshimi2002Nuclear,Bloom1962Principles, Chalupczak2015Alkali, Bevington2021Object}.
 These systems bifurcate at a critical feedback amplification factor $\alpha_c$: below $\alpha_c$, the spins relax to the equilibrium with a constant longitudinal component and zero transverse components of spin polarization; while above $\alpha_c$, the spin maser enters the limit cycle phase, persisting periodic oscillations in the transverse components of spin polarization. 
  Taking advantage of the precision frequency reference provided by limit cycles and coupling with magnetic fields, the spin masers are widely applied in various fields, including high-precision magnetometry~\cite{bevington2020magnetic,Gilles03,Jiang2021Floquet}, searches for permanent electric dipole moments~\cite{Rosenberry2001Atomic, Inoue2016Frequency}, tests of fundamental symmetry~\cite{Bear2000Limit, Safronova2018Search}, and probes of new physics~\cite{Afach2021Search, Terrano2022Comgnetometer, Safronova2018Search}. 
    
 However, such spin masers have no more dynamic phases, limiting the explorations therein. More recently, a system of coupled twin spin masers working in dual bias magnetic fields was studied~\cite{wang2023feedback} (see Fig.~\ref{fig: schematic}). In this system, there are much richer dynamic phases: besides limit cycles, quasi-periodic orbits and chaos also exist, extending the applications of spin masers. This also explained that the well-developed multi-species spin masers~\cite{Chupp1994Spin, Stoner1996Demonstration, Bear1998Improved, Sato2018Development, Bevington2020Dual} can be equivalently viewed as coupled spin masers operating in a special limit of quasi-periodic orbits.
 As the richer dynamic phases of the coupled twin spin masers identified~\cite{wang2023feedback}, an interesting question remains: what is the nature of the bifurcations at which the stable phases of the system turn unstable?
  
Since the coupled twin spin masers have six dimensions [see Eqs.~(\ref{blochx}) to (\ref{blochz})] which makes it hard to visualize trajectories of the system and thereby identify bifurcations, in this work, we turn to a three dimensional dynamic system closely related to the coupled twin spin masers [see Eqs.~(\ref{a}) to (\ref{pz})], and 
systematically study the bifurcations in the related system. As shown in the stability diagram Fig.~\ref{fig: stablediagram}, along different routes [labeled by \textcircled{\raisebox{-0.9pt}{1}}$\sim$\textcircled{\raisebox{-0.9pt}{5}}], the related system undergoes various types of bifurcations, including pitchfork, Hopf, homoclinic and saddle-node bifurcation of cycles.
  We reveal how these bifurcations are affected by the interplay between attractors, including various types of fixed points, both stable and unstable limit cycles etc., and evaluate the quantitative variations of these attractors, such as the positions of fixed points and the trajectories of limit cycles along different routes, which enables us to identify the locations of the corresponding bifurcations.
 Our findings suggest the nature of the bifurcations in the coupled twin spin masers and further elucidate the rich nonlinear phenomena in the system.
  
\section {Formalism} 
\begin{figure}
	\centering
	\includegraphics[width = 1 \linewidth]{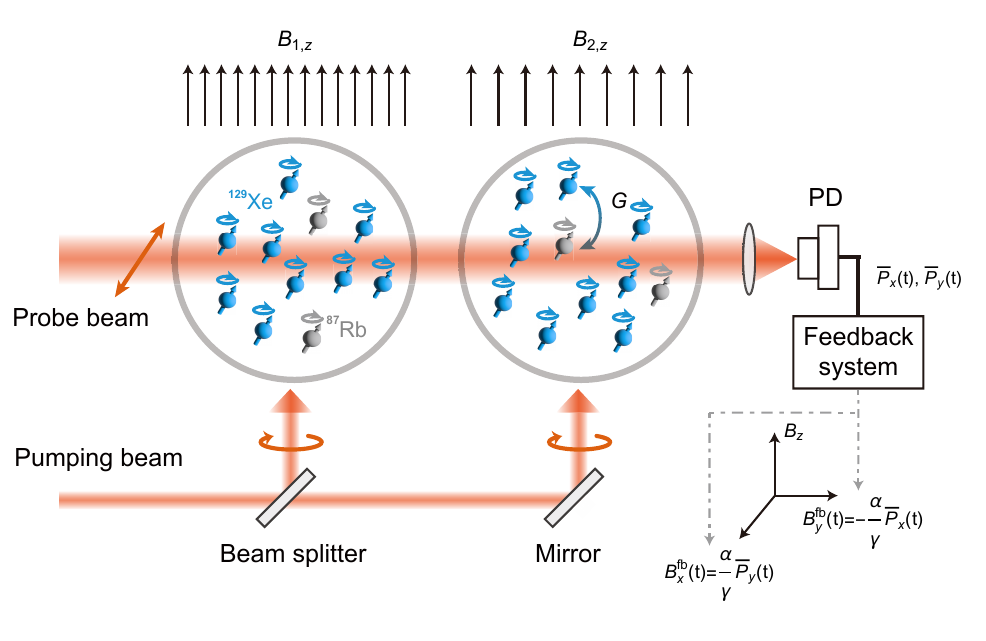}
	\caption{ Schematic of the feedback assisted twin spin masers setup. 
	Two identical cells, each containing noble gas $^{129}\rm Xe$ nuclear spins and  alkali-metal $^{87}\rm Rb$ atoms, are placed in two different bias magnetic fields $B_{1,z}$ and $B_{2,z}$ respectively, resulting in two intrinsic Larmor frequencies, $\omega_1$ and $\omega_2$.
	The alkali-metal $^{87}\rm Rb$ atoms, optically pumped by circularly polarized light, transfer polarization to the $^{129}\rm Xe$ nuclear spins through spin-exchange collisions.
	The twin spin masers are coupled by a feedback system, where the overall mean transverse polarization ($\overline{P}_x, \overline{P}_y$), detected by the photo diode (PD) monitoring the optical rotation of a linearly polarized probe beam, is simultaneously fed to produce the transverse magnetic field in the form of $B_x^{\text{fb}}(t)=\frac{\alpha}{\gamma}\overline P_{y}(t)$ and $B_y^{\text{fb}}(t)=-\frac{\alpha}{\gamma}\overline P_{x}(t)$ applied to both cells.}
	\label{fig: schematic}
\end{figure}

We consider two coupled spin masers placed in external magnetic fields as illustrated in Fig.~\ref{fig: schematic}. 
Each spin maser consists of a cell containing a $^{129}\rm Xe$ noble gas as the working spins and auxiliary $^{87}\rm Rb$ atoms. The $j$th maser is subject to an external bias field $\mathbf B_j=B_{j,z}\hat z$, which results in an intrinsic Larmor frequency for the $^{129}\rm Xe$ atoms, $\omega_j=\gamma B_{j,z}$, with $\gamma$ the nuclear gyromagnetic ratio.
The auxiliary $^{87}\rm Rb$ atoms serve two roles. On the one hand, the $^{87}\rm Rb$ atoms can be optically pumped and polarize the $^{129}\rm Xe$ nuclear spins through the spin-exchange collisions~\cite{Happer1972Optical,Happer1984Polarization,Walker1997Spin}. On the other hand, the $^{87}\rm Rb$ atoms can be employed to monitor the average spin polarization of $^{129}\rm Xe$ via the optical rotation of a linearly polarized probe beam detectable by the photo diode (PD) \cite{Jiang2021Floquet,Yoshimi2002Nuclear,Sato2018Development}. 
The two spin masers are coupled through the feedback fields $\mathbf B^{\text{fb}}=(B_x^{\text{fb}}(t),B_y^{\text{fb}}(t),0)=(\frac{\alpha}{\gamma}\overline P_{y}(t), -\frac{\alpha}{\gamma}\overline P_{x}(t),0)$, where $\overline{\mathbf P}(t)\equiv[{\mathbf P}_1(t)+{\mathbf P}_2(t)]/2$ is the overall mean spin polarization, $\mathbf P_j=(P_{j,x},P_{j,y},P_{j,z})$ is the average polarization of $^{129}\rm Xe$ nuclear spins in the $j$th cell, and $\alpha$ is the feedback amplification factor. The fields $\mathbf B^{\text{fb}}$ are generated by the feedback system which takes in the information of $\overline{\mathbf P}(t)$ through the signal from the photo diode.

The dynamics of $\{\mathbf P_1(t),\mathbf P_2(t)\}$ in such a spin system is governed by the following nonlinear Bloch equations ~\cite{wang2023feedback}
\begin{align}
\frac{d P_{j,x}}{dt}=&\omega_j P_{j,y}+\alpha \overline{P}_x P_{j,z}-\frac{P_{j,x}}{T_2 },\label{blochx}\\
\frac{d P_{j,y}}{dt}=&-\omega_j P_{j,x}+\alpha \overline{P}_y P_{j,z}-\frac{P_{j,y}}{T_2 },\label{blochy}\\
\frac{d P_{j,z}}{dt}=&-\alpha\left(\overline{P}_x P_{j,x}+\overline{P}_y P_{j,y}\right)-\frac{P_{j,z}}{T_1}+G(P_0-P_{j,z}).
\label{blochz}
\end{align}
The last term in Eq.~(\ref{blochz}) describes the pumping effect with $G$ the spin-exchange rate between the $^{129}\rm Xe$ and $^{87}\rm Rb$ atoms, and with $P_0$ the polarization of $^{87}\rm Rb$ atoms. The longitudinal and transverse spin relaxation times are $T_1$ and $T_2$ respectively. It has been shown that 
such a nonlinear system exhibits rich dynamic phases, including limit cycles, quasi-periodic orbits and chaos~\cite{wang2023feedback}. However, what is the nature of the transitions between these phases when parameters varied awaits being answered.


To shed light on the above question, we turn to another set of equations
\begin{align}
\frac{d A}{dt}=&\alpha \overline P_z A+\epsilon B/2-A/T_2 ,\label{a}\\
\frac{d B}{dt}=&-\epsilon A/2-B/T_2 ,\label{b}\\
\frac{d \overline P_z}{dt}=&-\alpha A^2/4-\overline P_z/T_1+G(P_0-\overline P_z)\label{pz},
\end{align}
with $\epsilon\equiv\omega_1-\omega_2$. This system, Eqs.~(\ref{a}) to (\ref{pz}), is closely related to Eqs.~(\ref{blochx}) to (\ref{blochz}): in most cases, the numerical solution to Eqs.~(\ref{blochx}) to (\ref{blochz}) has been shown to have the properties $|P_{1,T}|=|P_{2,T}|$ and $P_{1,z}=P_{2,z}$, meaning $\overline P_{T}\equiv (P_{1,T}+P_{2,T})/2$ is orthogonal to $\Delta P_T\equiv P_{1,T}-P_{2,T}$ with $P_{j,T}\equiv P_{j,x}+iP_{j,y}$~\cite{wang2023feedback}. These properties prompt one to re-parameterize $ A\equiv 2\overline P_T e^{-i\theta}$, $B \equiv \Delta P_T e^{-i(\theta+\pi/2)}$ and $\overline P_z\equiv P_{1,z}=P_{2,z}$. Making use of the above two properties and substituting the re-parameterization into Eqs.~(\ref{blochx}) to (\ref{blochz}), one can derive Eqs.~(\ref{a}) to (\ref{pz}), together with $\theta(t)=-\omega_c t+\phi$ where $\omega_c=(\omega_1+\omega_2)/2$ and $\phi$ is an arbitrary phase. 

\begin{figure}
	\centering
	\includegraphics[width = 1 \linewidth]{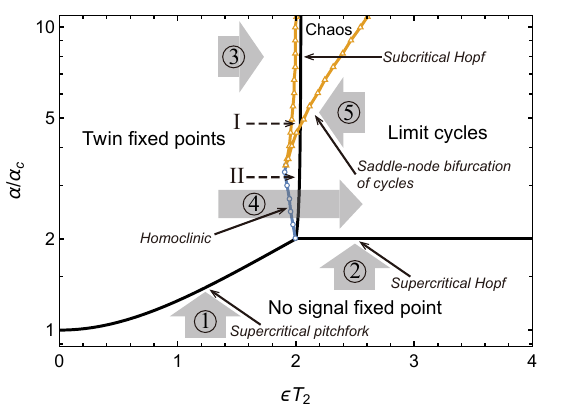}
	\caption{Stability diagram of the dimension reduced system, Eqs.~(\ref{a}) to (\ref{pz}). The stable dynamic phases include the twin fixed points, the limit cycles, the chaos and the no signal fixed point.
	The black solid curves plot Eqs.~(\ref{eq: stableBoundary1}) and~(\ref{eq: stableBoundary2}), marking the stable region boundaries of the no signal fixed point and the twin fixed points. The line-linked symbols of triangles and circles mark the stable region boundaries of the chaos and the limit cycles which are determined numerically by taking $P_0=1$, $T_1=21.5 \rm s$, $T_2=13.65 \rm s$, $G=0.03 \rm Hz$. There exist Region I co-stable for the twin fixed points and the chaos, and Region II for the twin fixed points and the limit cycles. The labels \textcircled{\raisebox{-0.9pt}{1}}$\sim$\textcircled{\raisebox{-0.9pt}{5}}, together with the large gray arrows, mark the routes along which different types of bifurcations we study. Various types of bifurcations are pointed out at the boundaries.}
	\label{fig: stablediagram}
\end{figure}

As shown in Fig.~\ref{fig: stablediagram}, the dimension reduced system, Eqs.~(\ref{a}) to (\ref{pz}), exhibits three nontrivial phases: twin fixed points, limit cycles and chaos.  The trivial no signal fixed point has 
\begin{align}
A_{\rm NS}=& B_{\rm NS}=0, \label{ans}\\ 
\overline P_{z,\rm NS}=&P_0/(1+1/GT_1).\label{pns}
\end{align}
 For the phase of twin fixed points, we find 
 \begin{align}
 A_{\rm TFP,\pm}=&\pm 2\{GP_0[\alpha-\alpha_cf(\epsilon T_2)]\}^{1/2}/\alpha\label{atfp}\\
 B_{\rm TFP,\pm}=&-\epsilon T_2 A_{\rm TFP,\pm}/2\\
 \overline P_{z,{\rm TFP}}=&f(\epsilon T_2)/\alpha T_2\label{ptfp}
 \end{align} 
 with $\alpha_c=(1+1/GT_1)/T_2P_0$ the critical amplification factor for $\epsilon=0$ and $f(x)=1+(x/2)^2$ ($\pm$ corresponds to one of the twin fixed points).
These three nontrivial phases of Eqs.~(\ref{a}) to (\ref{pz}) correspond to the limit cycle, quasi-periodic orbit and chaotic phases of  Eqs.~(\ref{blochx}) to (\ref{blochz}) respectively~\cite{wang2023feedback} (also see Appendix A). 
 A notable advantage of working with the three dimensional system, Eqs.~(\ref{a}) to (\ref{pz}), other than the full six dimensional system, Eqs.~(\ref{blochx}) to (\ref{blochz}), is that the three dimensional trajectories of $\{A, B, \overline{P}_z\}$ can be conveniently visualized. Thus, hereafter we focus on studying the bifurcations of the dimension reduced system, Eqs.~(\ref{a}) to (\ref{pz}).

\section{Bifurcations}
Figure~\ref{fig: stablediagram} is the stability diagram of Eqs.~(\ref{a}) to (\ref{pz}) taking parameters $P_0=1$, $T_1=21.5 \rm s$, $T_2=13.65 \rm s$, $G=0.03 \rm Hz$ \cite{Jiang2021Floquet}.
The stable region boundaries of the no signal fixed point and twin fixed point phases, represented by the solid lines, can be derived analytically through linear stability analysis~\cite{wang2023feedback}. The upper boundary for the no signal fixed point phase is given by 
\begin{align}
	 \alpha/\alpha_c &= \begin{cases}1 + (\epsilon T_2 /2)^2 
  &  \text{ for } \epsilon T_2  < 2; \\
  2 & \text{ for } \epsilon T_2  \geq 2.
\end{cases}\label{eq: stableBoundary1}
\end{align}
The right boundary for the twin fixed point phase is 
\begin{align}
\alpha/\alpha_c &= \frac{3}{2}y +\frac{1-d}{2(y-d)}\label{eq: stableBoundary2}
\end{align}
for $\alpha/\alpha_c>2$, where $y=(\epsilon T_2 /2)^2$, $d\equiv T_2 (G+1/T_1)$, and $\min(1,d)<y<\max(1,d)$.
The rest stable region boundaries for the limit cycles and chaos phases, marked by the hollow triangle and circle symbols, are determined numerically. In Region I, the chaos and twin fixed point phases are both stable, whereas the limit cycles and twin fixed point phases in Region II. In these two regions, the eventual long time stable dynamic behaviors depend on the initial conditions. When the system parameters varied, the stable dynamics of the system can cross different boundaries along the routes \textcircled{\raisebox{-0.9pt}{1}}$\sim$\textcircled{\raisebox{-0.9pt}{5}}, as labeled in Fig.~\ref{fig: stablediagram}. 
In the following, we are going to show that at the boundary-crossings, the system undergoes bifurcations of various types including \emph{pitchfork, Hopf, homoclinic}, and \emph{saddle-node bifurcation of cycles}.


\begin{figure}
	\centering
	\includegraphics[width =1\linewidth]{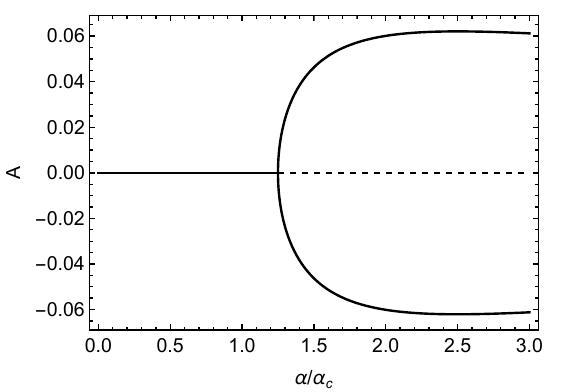}
	\caption{Illustration of the pitchfork bifurcation on Route \textcircled{\raisebox{-0.9pt}{1}} with $\epsilon T_2=1$. For $\alpha/\alpha_c<1+(\epsilon T_2/2)^2$, the no signal fixed point with $A=0$ is stable, represented by the solid line. Beyond the critical point at $\alpha/\alpha_c=1+(\epsilon T_2/2)^2$, the no signal fixed point with $A=0$ is unstable, represented by the dashed line, instead the stable twin fixed points emerges, whose $A$-component are non zero, represented by the two branching solid lines.}
	\label{fig: Pitchfork}
\end{figure}

\subsection{Pitchfork bifurcation}

Pitchfork bifurcations typically involve the splitting of fixed points in parameter space, and can happen either \emph{supercritically} or \emph{subcritically}~\cite{Strogatz2018Nonlinear}. For a supercritical bifurcation, the stable fixed point splits into new fixed points without lossing stability as the parameter varies; otherwise, it indicates a subcritical one.

In our system, 
as $\alpha$ increases along Route \textcircled{\raisebox{-0.9pt}{1}} with fixed $\epsilon$ (for $\epsilon T_2<2$), the no signal fixed point turns to be unstable at the point $\alpha/\alpha_c=1+(\epsilon T_2/2)^2$. The transition is due to a \emph{supercritical} pitchfork bifurcation: in the phase space of $\{A,B,\overline P_z\}$, a pair of twin fixed points $(A_{\rm TFP,\pm}, B_{\rm TFP,\pm}, \overline{P}_{z, \rm TFP})$ emerge at the exact location of the no signal fixed point  $(A_\mathrm{NS}, B_\mathrm{NS},\overline{P}_{z,\mathrm{NS}})$ [see Eqs.~(\ref{ans}) and (\ref{ptfp})]. When $\alpha$ continues increasing beyond this critical point, the phase of twin fixed points becomes stable, and $|A_{\rm TFP,\pm}|$ and $|B_{\rm TFP,\pm}|$ move away from zero.
Figure~\ref{fig: Pitchfork} illustrates the pitchfork bifurcation, providing an immediate understanding of why this term is used~\cite{Strogatz2018Nonlinear}.

\begin{figure}
	\centering
	\includegraphics[width= 0.85\linewidth]{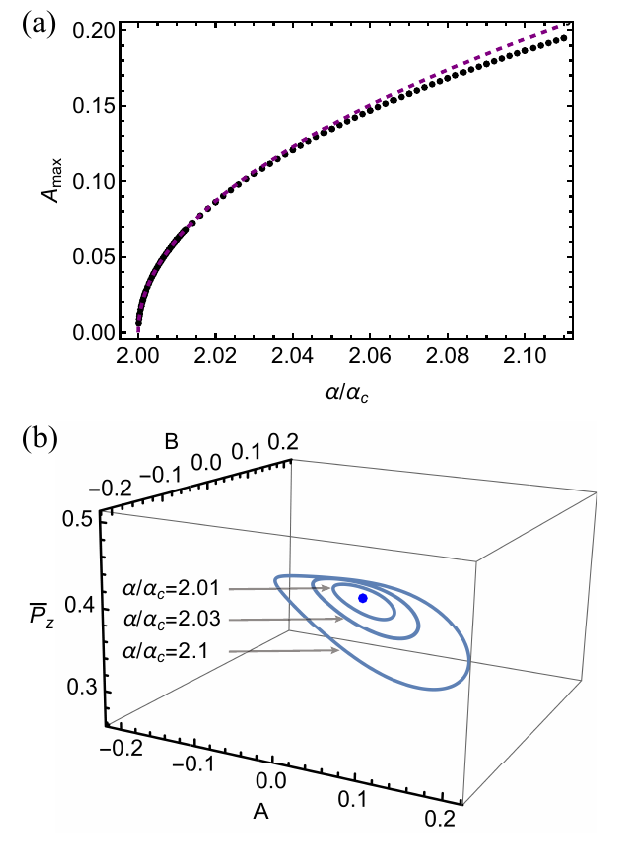}
	\caption{Illustration of the supercritical Hopf bifurcation near $\alpha/\alpha_c=2$ on Route \textcircled{\raisebox{-0.9pt}{2}} with $\epsilon T_2=2.5$. (a) The $A$-component maxima of the stable limit cycle trajectory $A_{\rm max}$ versus $\alpha/\alpha_c$. The black dots are obtained by numerically solving Eqs.~(\ref{a})$\sim$(\ref{pz}); the dashed purple curve is the analytic perturbation result from Eq.~(\ref{eq: landalpha}). (b) The blue curves are the trajectories of the limit cycles for different $\alpha$. The blue dot therein represents the unstable no signal fixed point for reference.}
	\label{fig: HopfForepsilon2point5}
\end{figure}

\subsection{Hopf bifurcation}

Fixed points can also lose stability through Hopf bifurcations. Similar to pitchfork bifurcations, there are also two types of Hopf bifurcations: \emph{supercritical} and \emph{subcritical}. If a stable, small-amplitude limit cycle emerges at the location of a fixed point after crossing a bifurcation point, it is called a \emph{supercritical} Hopf bifurcation; if a limit cycle shrinks into zero dimension and disappears at a bifurcation point, 
it is called a \emph{subcritical} Hopf bifurcation~\cite{Strogatz2018Nonlinear}.
We find that our dimension reduced system exhibits both types of Hopf bifurcations.

First, as $\alpha$ increases along Route \textcircled{\raisebox{-0.9pt}{2}} with fixed $\epsilon$ (for $\epsilon T_2>2$), the system undergoes a \emph{supercritical} Hopf bifurcation at $\alpha/\alpha_c=2$. This can be seen from the eigenvalues of the Jocobian at the no signal fixed point \cite{wang2023feedback}, which reads
\begin{align}
	\lambda_{\mathrm{NS},\pm}=\frac{1}{2T_2}\left\{\alpha/\alpha_c-2\pm\left[(\alpha/\alpha_c)^2-(\epsilon T_2)^2\right]^{1/2}\right\}.
\end{align}
As $\alpha$ increases, in the complex plane, the two separate $\lambda_{\mathrm{NS},\pm}$ simultaneously cross the imaginary axis from the left to the right; namely $\rm Re(\lambda_{\mathrm{NS},\pm})=0$ while the pair of  $\rm Im(\lambda_{\mathrm{NS},\pm})$ are nonzero at the transition point $\alpha/\alpha_c=2$, which is typical to a Hopf bifurcation.
At the same time, in the phase space of $\{A,B,\overline P_z\}$, a stable limit cycle emerges at the location of the no signal fixed point when $\alpha$ exceeds $2 \alpha_c$, which confirms that the Hopf bifurcation here is a \emph{supercritical} one.
Figure~\ref{fig: HopfForepsilon2point5} (b) shows how the limit cycles vary with $\alpha$. To quantitatively illustrate this process, we calculate $A_{\rm max}$ (the maxima of $A$-component of the trajectory of the limit cycles) by numerically solving the dynamic equations~(\ref{a})$\sim$(\ref{pz}). As shown in Fig.~\ref{fig: HopfForepsilon2point5}(a), $A_{\rm max}\rightarrow 0$ as $\alpha/\alpha_c\rightarrow 2$, indicating the coincidence between the limit cycles and the no signal fixed point. The numerical results of $A_{\rm max}$ around $\alpha/\alpha_c=2$ can be largely reproduced by an analytic perturbation calculation given in Appendix B, which also shows that, in the limit $\alpha/\alpha_c\to 2$, the trajectory size of the limit cycles $l\varpropto\sqrt{\alpha/\alpha_c-2}$ and the period of the limit cycles $\tau\to 2\pi/\mathrm{Im}(\lambda_{\mathrm{NS},+})$. 
Both of these asymptotic behaviors agree with a \emph{supercritical} Hopf bifurcation~\cite{Strogatz2018Nonlinear}.

\begin{figure}[htbp]
	\centering
	\includegraphics[width=0.85\linewidth]{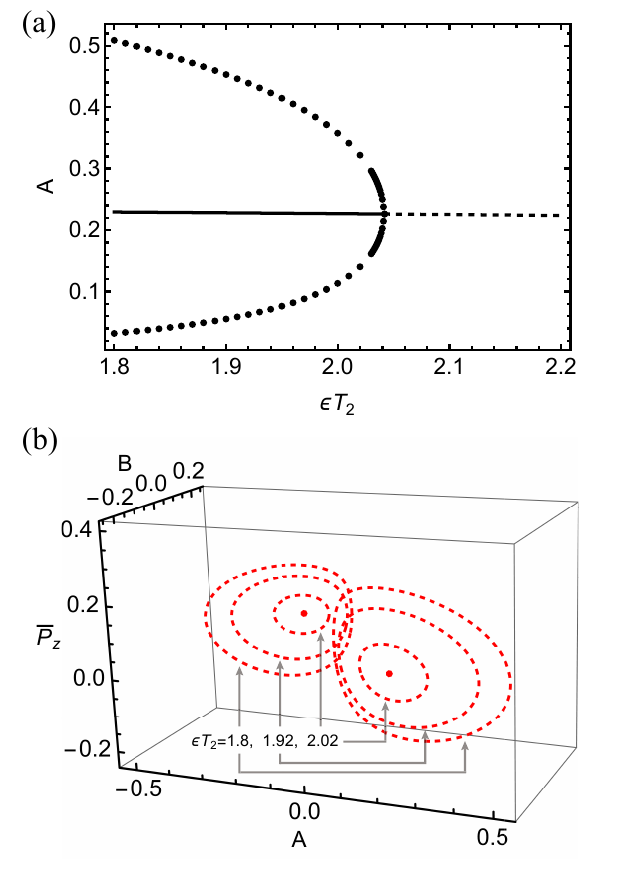}
	\caption{Illustration of the subcritical Hopf bifurcation on Route \textcircled{\raisebox{-0.9pt}{3}} with $\alpha/\alpha_c=10$. (a) The black dots are the maximum and minimum $A$-component of the pair of unstable limit cycles; the solid (dashed) line denotes the $A$-component of the stable (unstable) twin fixed points. Here only the $A>0$ part is shown for simplicity due to the invariance of the system under $A\mapsto-A$ and $B\mapsto-B$.
	(b) The red dashed curves are the trajectories of the unstable limit cycle pair for different $\epsilon T_2$; the red dots denote the twin fixed points at $\epsilon=\epsilon_{\rm Hopf}$.}
	\label{HomoclinicForalpha10}
\end{figure}

Second, as shown in Fig.~\ref{fig: stablediagram}, the twin fixed points become unstable at $\alpha/\alpha_c = 3y/2+(1-d)/2(y-d)$ with $\epsilon$ increasing. Along Route \textcircled{\raisebox{-0.9pt}{3}} or \textcircled{\raisebox{-0.9pt}{4}}, after passing the overlapped Region I or (and) II, only chaos or the limit cycles continue to remain stable in the right vicinity of this critical line, which indicates a \emph{subcritical} Hopf bifurcation.

\begin{figure*}[htbp]
	\centering
	\includegraphics[width=0.8\textwidth]{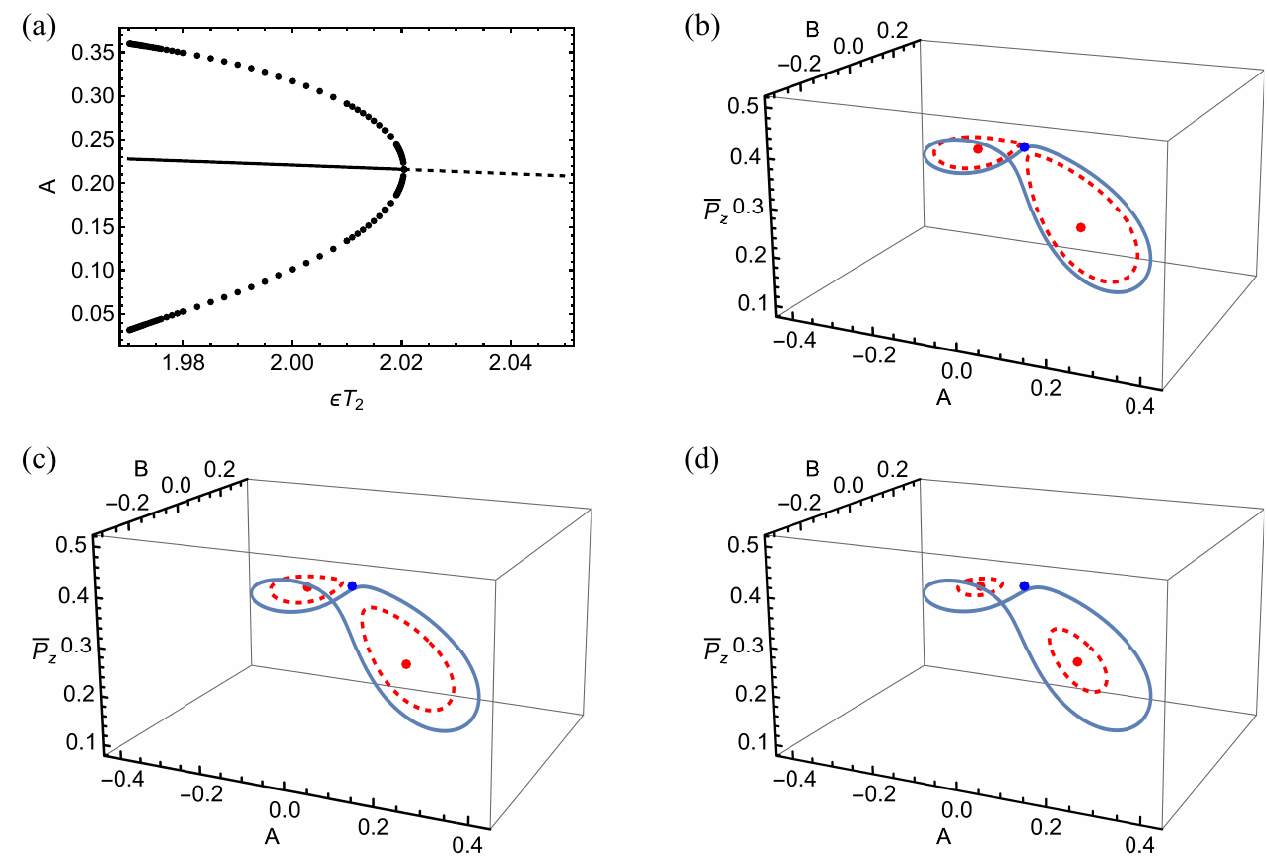}
	\caption{Illustration of the subcritical Hopf bifurcation and the homoclinic bifurcation on Route \textcircled{\raisebox{-0.9pt}{4}} with $\alpha/\alpha_c=2.46$. (a) The black dots are the maximum and minimum $A$-component of the pair of unstable limit cycles; the solid (dashed) line denote the $A$-component of the stable (unstable) twin fixed points. 
 (b)$\sim$(d) show the trajectories of the stable limit cycles (the blue curves) and the unstable limit cycle pairs (the red dashed curves) with $\epsilon T_2=1.97$, $\epsilon T_2=1.99$ and $\epsilon T_2=2.01$ respectively. The blue and red dots denote the no signal fixed point and the twin fixed points respectively.}
	\label{HomoclinicForalpha2point46}
\end{figure*}

Along Route \textcircled{\raisebox{-0.9pt}{3}}, we take $\alpha/\alpha_c=10$ for an instance.
Figure~\ref{HomoclinicForalpha10} (b) shows that there exist a pair of unstable limit cycles, each of which encloses one of the twin fixed points when $\epsilon$ is taken to be in the left vicinity of the curve $\alpha/\alpha_c = 3y/2+(1-d)/2(y-d)$. Figure~\ref{HomoclinicForalpha10}(a) plots the maximum and minimum $A$-component of the unstable limit cycle pair; it can be seen that as $\epsilon$ approaches the critical line, the unstable limit cycle pair gradually shrink until they coincide with the twin fixed points at $\epsilon=\epsilon_{\rm Hopf}$ where a Hopf bifurcation occurs, rendering the twin fixed points unstable. Along Route \textcircled{\raisebox{-0.9pt}{4}}, we take $\alpha/\alpha_c=2.46$ to illustrate the bifurcation. Figure~\ref{HomoclinicForalpha2point46} shows that the situation is similar to the above case for $\alpha/\alpha_c=10$.

 
\begin{figure}
	\centering
\includegraphics[width=0.85\linewidth]{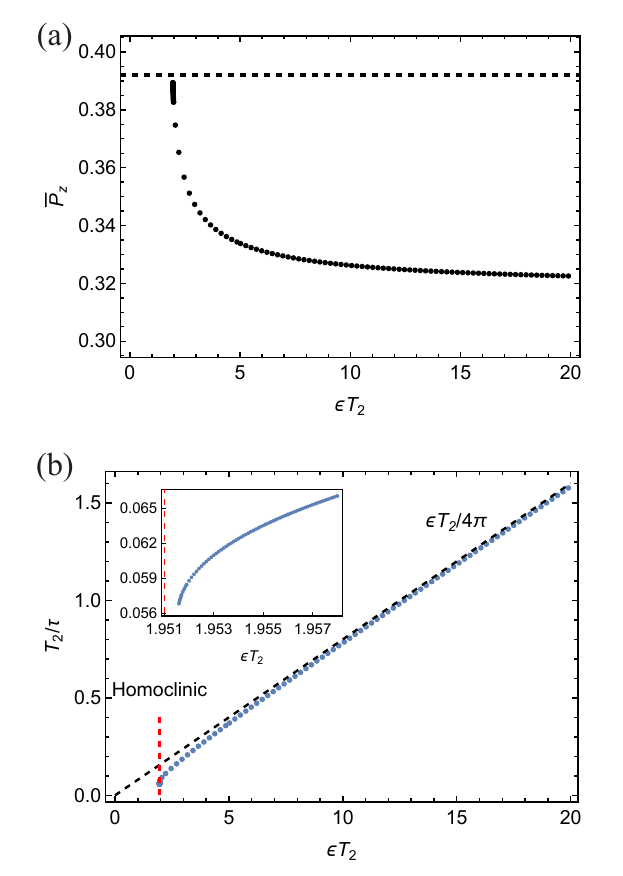}
	\caption{Illustration of the homoclinic bifurcation regarding the limit cycles on Route \textcircled{\raisebox{-0.9pt}{4}} with $\alpha/\alpha_c=2.46$. (a) The maxima of $\overline P_{z}$ of the limit cycles (the black dots) versus $\epsilon$. The black dashed line is $\overline P_{z,\rm NS}$ representing the no signal fixed point. (b) The period $\tau$ of the limit cycles (the blue dots) versus $\epsilon$. The black dashed line is the asymptote $1/\tau=\epsilon/4\pi$ which is anticipated for $\epsilon\rightarrow\infty$. The red dashed line marks the location of the homoclinic bifurcation, with the nearby details shown in the inset.}
	\label{fig: homoLC}
\end{figure}

\subsection{Homoclinic bifurcation}
A homoclinic bifurcation occurs when a limit cycle intersects with an unstable fixed point, typically a saddle point~\cite{Strogatz2018Nonlinear}. In the reverse direction along Route \textcircled{\raisebox{-0.9pt}{4}}, when approaching the left boundary of the limit cycle phase, Fig.~\ref{HomoclinicForalpha2point46}(b)$\sim$(d) show that the trajectory of the limit cycles gradually deforms and touches the \emph{unstable} no signal fixed point at the boundary; a homoclinic bifurcation occurs and the stable limit cycle is annihilated afterwards. 
Figure~\ref{fig: homoLC}(a) shows that the maximum $\overline{P}_{z}$ of the limit cycle trajectory (the black dots) gradually moves close to $\overline{P}_{z,\rm NS}$ until the limit cycles pass through the no signal fixed point, leading to the homoclinic bifurcation. 
Figure~\ref{fig: homoLC}(b) shows the period $\tau$ of the limit cycle diverges asymptotically as $\epsilon$ decreases toward the bifurcation point, which is a key feature of the homoclinic bifurcation.



\begin{figure*}[htbp]
	\centering
	\includegraphics[width=0.8\textwidth]{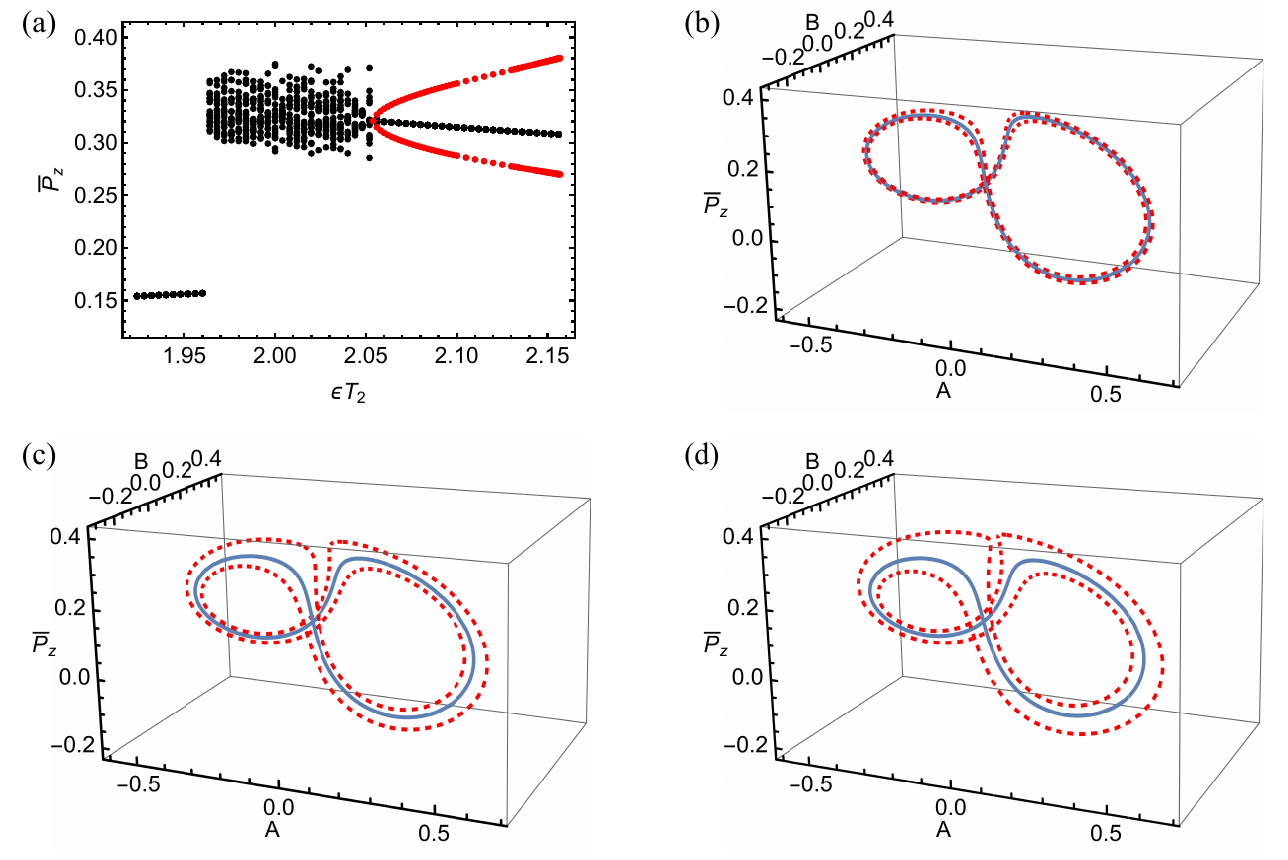}
	\caption{Illustration of the saddle-node bifurcation of cycles on Route \textcircled{\raisebox{-0.9pt}{5}} with $\alpha/\alpha_c=4.9$.
	(a) The red dots are the maxima of $\overline{P}_z$, denoted by $\overline{P}_{z,\rm max}$, of the unstable limit cycles [represented by the red dashed lines in (b)$\sim$(d)]. The black dots correspond to $\overline{P}_{z,\rm max}$ in the steady dynamics within a duration of $1000\text{s}$. The random scattering of the maximum $\overline{P}_z$ indicates chaos, whereas the smooth $\overline{P}_{z,\rm max}$ represents the stable limit cycles (right to the chaos) or the twin fixed points (left to the chaos).
	The saddle-node bifurcation of cycles happens where the pair of unstable limit cycles coincide with the stable ones. (b)$\sim$(d) show the trajectories of the unstable (the red dashed curves) and the stable (the blue solid curve) limit cycles for  $\epsilon T_2=2.05$, $\epsilon T_2=2.11$ and $\epsilon T_2=2.16$ respectively.}
	\label{HomoclinicForalpha4point9}
\end{figure*}

\subsection{Saddle-node bifurcation of cycles}
Along Route \textcircled{\raisebox{-0.9pt}{5}} in Fig.~\ref{fig: stablediagram}, when hitting the right boundary of the chaos phase, the limit cycles become unstable, and the system undergoes a saddle-node bifurcation of cycles. To illustrate this bifurcation, we take $\alpha/\alpha_c=4.9$ with varying $\epsilon$. This transition occurs because when $\epsilon$ is slightly above its critical value, in addition to the stable limit cycles, there exist a pair of unstable limit cycles. Figure~\ref{HomoclinicForalpha4point9} shows that as $\epsilon$ decreases, the trajectories of the unstable limit cycles gradually approach the one of the stable, and eventually they coincide at the transition. This is a saddle-node bifurcation of cycles~\cite{Strogatz2018Nonlinear}, leading to the disappearance of the stable limit cycles and the emergence of chaos.
In Fig.~\ref{HomoclinicForalpha4point9}(a), the black dots represent the local maxima of $\overline{P}_z$, namely $\overline{P}_{z,\rm max}$, extracted from a stable trajectory of duration $1000\rm s$. In the chaos phase, $\overline{P}_{z,\rm max}$ scatters around due to the irregular wandering of the trajectory in the phase space, while the motion of the stable limit cycles leads to smooth $\overline{P}_{z,\rm max}$ as a function of $\epsilon$. 


\section{Conclusion}
To shed light on the bifurcations of nonlinear dynamics in coupled twin spin masers placed in dual bias magnetic fields, 
we have systematically studied the dimension reduced system, Eqs.~(\ref{a})$\sim$(\ref{pz}), whose phases correspond to those of the former one by one (see Fig.~\ref{fig: stablediagram} and Appendix A). We have identified the different types of bifurcations in the dimension reduced system, and summarize the results along the different routes below.

(i) Along Routes \textcircled{\raisebox{-0.9pt}{1}} and \textcircled{\raisebox{-0.9pt}{2}}. As $\alpha$ increases, the no signal fixed point splits into a pair of stable twin fixed points at the transition line  $\alpha/\alpha_c=1+(\epsilon T_2)^2/2$ (for $\epsilon T_2<2$), leading to a \emph{supercritical} pitchfork bifurcation on Route \textcircled{\raisebox{-0.9pt}{1}}. For $\epsilon T_2>2$, when $\alpha$ exceeds $2 \alpha_c$, the no signal fixed point loses stability to the limit cycles whose trajectory, in the phase space, emerges at the position of the no signal fixed point, and grows in size as $l\varpropto\sqrt{\alpha/\alpha_c-2}$. This bifurcation on Route \textcircled{\raisebox{-0.9pt}{2}} is of the \emph{supercritical} Hopf type.

(ii) Along Routes \textcircled{\raisebox{-0.9pt}{3}} and \textcircled{\raisebox{-0.9pt}{4}}. To the left of the line $\alpha/\alpha_c = 3y/2 + (1-d)/2(y-d)$, there exist a pair of unstable limit cycles coexisting with the stable twin fixed points.  As $\epsilon$ increases, the trajectories of the unstable limit cycle pair gradually shrink till they coincide with that of the twin fixed points, leading the latter to lose stability. After this \emph{subcritical} Hopf bifurcation, either the chaos on Route \textcircled{\raisebox{-0.9pt}{3}} or the limit cycles on Route \textcircled{\raisebox{-0.9pt}{4}} remain the sole stable dynamics of the system. Conversely, as $\epsilon$ decreases on Route \textcircled{\raisebox{-0.9pt}{4}}, there is a homoclinic bifurcation where the initially stable limit cycles touch the unstable no signal fixed point and become unstable.

(iii) Along Route \textcircled{\raisebox{-0.9pt}{5}}. To the right of the line $\alpha/\alpha_c = 3y/2 + (1-d)/2(y-d)$, there exists a saddle-node bifurcation of cycles where a pair of unstable limit cycles coincide with the initially stable one. 

The pitchfork, Hopf, homoclinic bifurcations and saddle-node bifurcation of cycles diagnosed in the reduced dimension system, Eqs.~(\ref{a}) to (\ref{pz}), are likely to be the types of the bifurcations occurring in the full system, Eqs.~(\ref{blochx}) to (\ref{blochz}), given the close relation between the two systems. 
Our findings deepen the understanding of the underlying mechanisms resulting in the rich dynamic phases in the coupled twin spin masers.
Previously it has been shown that the spin dynamics of coupled multiple spin masers other than two would exhibit qualitatively the same nonlinear phases as Eqs.~(\ref{a}) to (\ref{pz}) ~\cite{wang2023feedback}. It is natural to expect that the types of bifurcations found here are generic in such a rather wide class of spin systems.


\section{Acknowledgements}
We thank Xiaodong Li and Long Wang for their discussions on numerics. TW and ZY are supported by the National Natural Science Foundation of China Grant No.12074440, and Guangdong Project (Grant No.~2017GC010613). 

\section*{Appendix}
\subsection{ Correspondence between the phases of the two systems }
Via the reparameterization $ A\equiv 2\overline P_T e^{-i\theta}$, $B \equiv \Delta P_T e^{-i(\theta+\pi/2)}$ and $\overline P_z\equiv P_{1,z}=P_{2,z}$, one can derive Eqs.~(\ref{a}) to (\ref{pz}) from Eqs.~(\ref{blochx}) to (\ref{blochz}). Thus the stable trajectories obtained for the former in terms of $\{A,B,\overline P_z\}$ can be mapped back to those in terms of $ \{\mathbf P_1,\mathbf P_2\}$. 
It is obvious that the trajectories of the no signal fixed point and chaos, once transformed to those in terms of $ \{\mathbf P_1,\mathbf P_2\}$, shall also be identified as the no signal fixed point and chaos respectively. 
For the twin fixed points, 
\begin{align}
	&P_{j,x}=\frac{|A_{\rm TFP,\pm}|}{2}[(-1)^j\frac{\epsilon T_2}{2}\sin(\omega_ct-\phi)+\cos(\omega_ct-\phi)],\label{eq: LCPx}\\
	&P_{j,y}=\frac{|A_{\rm TFP,\pm}|}{2}[(-1)^j\frac{\epsilon T_2}{2}\cos(\omega_ct-\phi)-\sin(\omega_ct-\phi)],\label{eq: LCPy}\\
	&P_{1,z}=P_{2,z}=\overline P_{z,{\rm TFP}},\label{eq: PzLC}
\end{align}
which shall be identified as limit cycles. 
The limit cycle phase of Eqs.~(\ref{a}) to (\ref{pz}) has the properties $A(t)=A(t+\tau)$, $B(t)=B(t+\tau)$ and $\overline P_z(t)=\overline P_z(t+\tau)$ with nonzero period $\tau$. Correspondingly,
\begin{align}
	&P_{j,x}(t)=\frac{A(t)}{2}\cos(\omega_ct-\phi)-(-1)^j\frac{B(t)}{2}\sin(\omega_ct-\phi),\label{eq: qpoPx}\\
	&P_{j,y}(t)=-\frac{A(t)}{2}\sin(\omega_ct-\phi)-(-1)^j\frac{B(t)}{2}\cos(\omega_ct-\phi).\label{eq: qpoPy}\end{align}
Since generally $\omega_c\tau/2\pi$ is not an integer, the trajectories represented in terms of $ \{\mathbf P_1,\mathbf P_2\}$ are quasi-periodic orbits. Therefore, qualitatively, any phase of  Eqs.~(\ref{blochx})$\sim$(\ref{blochz}) finds a corresponding one of Eqs.~(\ref{a})$\sim$(\ref{pz}).

\subsection{Perturbation calculation for the supercritical Hopf bifurcation}
\label{sec: perturbation}
Certain properties close to the supercritical Hopf bifurcation at $\alpha/\alpha_c=2$ can be otained analytically in a perturbative way.
We expand the limit cycle solution for $\alpha/\alpha_c\gtrsim2$ as $A=l\sin(\omega t)$ and  $\overline P_z=\overline P_{z,\rm NS}+\delta \overline P_z $, where $\omega=2\pi/\tau$ with $\tau$ the period of the limit cycles. Substituting the above ansatz into Eqs.~(\ref{b})$\sim$(\ref{pz}), we have
\begin{align}
	B&=-\frac{l \epsilon  T_2}{2[(\omega T_2)^2+1]}[-\omega T_2 \cos(\omega t)+ \sin(\omega t)],\label{eq: B}\\
	\delta \overline P_z&=-\frac{\alpha  l^2}{8}\left[\frac{1}{G+1/T_1}\right.\nonumber\\
	&\left.-\frac{(G+1/T_1)\cos(2\omega t)+2\omega\sin(2\omega t)}{(G+1/T_1)^2+4\omega^2}\right].	\label{eq: deltaPz}
\end{align}
Finally we substitute $A=l\sin(\omega t)$ and Eqs.~(\ref{eq: B}) to (\ref{eq: deltaPz}) into Eq.~(\ref{a}), 
and, by requiring the coefficients of $\sin\omega t$ and $\cos\omega t$ being zero, obtain
\begin{widetext}
\begin{align}
	\frac{\alpha}{\alpha_c}-\frac{l^2(G+1/T_1)}{16(GP_0)^2T_2}\left[\frac{3(GT_2+T_2/T_1)^2+8(\omega T_2)^2}{(GT_2+T_2/T_1)^2+4(\omega T_2)^2}\right]\left(	\frac{\alpha}{\alpha_c}\right)^2-\frac{(\epsilon T_2)^2}{4[1+(\omega T_2)^2]}&=1,\label{eq: eqns1}\\	
	\frac{l^2(G+1/T_1)^2}{8(GP_0)^2[(GT_2+T_2/T_1)^2+4(\omega T_2)^2]}\left(	\frac{\alpha}{\alpha_c}\right)^2+\frac{(\epsilon T_2)^2}{4[1+(\omega T_2)^2]}&=1.\label{eq: eqns2}
\end{align}

After eliminating $\omega$ from Eqs.~(\ref{eq: eqns1}) to (\ref{eq: eqns2}), we would have a relation between $l$ and $\alpha$ of the form $l^2=g(\alpha)$; keeping $g(\alpha)$ up to the first order of $\alpha/\alpha_c-2$, we derive
\begin{align}
		l=k\sqrt{\alpha/\alpha_c-2},\label{eq: landalpha}
\end{align}
with
	\begin{align}
		k=\sqrt{\frac{4G^2 P_0^2 T_1 T_2[T_2^2+2GT_1 T_2^2+T_1^2(-4+G^2 T_2^2+\epsilon^2 T_2^2)]}{(1+GT_1)\{3 T_2^2+2T_1 T_2(-1+3G T_2)+T_1^2[GT_2(-2+3GT_2)+2(-4+\epsilon^2T_2^2)]\}}}.
	\end{align}
\end{widetext}
The purple dashed curve in Fig.~\ref{fig: HopfForepsilon2point5}(a) plots the relation~(\ref{eq: landalpha}), which is consistent with the numerical results (the black dots). Note that the asymptotic behavior of $l\varpropto\sqrt{\alpha/\alpha_c-2}$ as $\alpha$ approaches $2\alpha_c$ is a critical feature of a \emph{supercritical} Hopf bifurcation. 

Similarly, after eliminating $l$ from Eqs.~(\ref{eq: eqns1})$\sim$(\ref{eq: eqns2}), we would obtain a relation between $\tau$ and $\alpha$. Notably, at $\alpha/\alpha_c=2$, $\tau=2\pi/\mathrm{Im}(\lambda_{\mathrm{NS},+})$~\cite{wang2023feedback}, another hallmark of a \emph{supercritical} Hopf bifurcation.

\bibliographystyle{apsrev4-1}
\bibliography{main}

\end{document}